\documentclass[pdflatex,sn-mathphys-num, i  icol]{sn-jnl}

\usepackage{graphicx}%
\usepackage{multirow}%
\usepackage{amsmath,amssymb,amsfonts}%
\usepackage{amsthm}%
\usepackage{mathrsfs}%
\usepackage[title]{appendix}%
\usepackage{xcolor}%
\usepackage{textcomp}%
\usepackage{manyfoot}%
\usepackage{booktabs}%
\usepackage{algorithm}%
\usepackage{algorithmicx}%
\usepackage{algpseudocode}%
\usepackage{listings}%
\usepackage{siunitx}
\usepackage{makecell}

\theoremstyle{thmstyleone}%
\theoremstyle{thmstyletwo}%

\theoremstyle{thmstylethree}%

\newcommand{\electron}{\ensuremath{\,{\mathrm{e^-}}\,}}
\newcommand{\posmu}{\ensuremath{\,{\mathrm{\mu^+}}\,}}

\newcommand{\keV}{\ensuremath{\,{\mathrm{keV}}\,}}

\newcommand{\MHz}{\ensuremath{\,{\mathrm{MHz}}\,}}

\def\nrtemplate#1#2#3{#1^{{\rm NR}#3}_{#2}}
\def\Vnrf#1#2{\nrtemplate{{{\mathcal V}_{#1}}}{#2}{}}
\def\TzBnrf#1#2{\nrtemplate{{{\mathcal T}_{#1}}}{#2}{(0B)}}
\def\ToBnrf#1#2{\nrtemplate{{{\mathcal T}_{#1}}}{#2}{(1B)}}
\def\anrf#1#2{\nrtemplate{{a_{#1}}}{#2}{}}
\def\cnrf#1#2{\nrtemplate{{c_{#1}}}{#2}{}}
\def\gzBnrf#1#2{\nrtemplate{{g_{#1}}}{#2}{(0B)}}
\def\goBnrf#1#2{\nrtemplate{{g_{#1}}}{#2}{(1B)}}
\def\HzBnrf#1#2{\nrtemplate{{H_{#1}}}{#2}{(0B)}}
\def\HoBnrf#1#2{\nrtemplate{{H_{#1}}}{#2}{(1B)}}
\def\sVnrf#1#2{\nrtemplate{{{\mathcal V}_{#1}}}{#2}{,{\rm Sun}}}

\def\sTzBnrf#1#2{\nrtemplate{{{\mathcal T}_{#1}}}{#2}{(0B),{\rm Sun}}}
\def\sToBnrf#1#2{\nrtemplate{{{\mathcal T}_{#1}}}{#2}{(1B),\rm Sun}}

\begin{document}

\title[Article Title]{Muonium fine structure: theory update, tests of Lorentz violation and experimental prospects}

\author*[1]{\fnm{Philipp} \sur{Blumer}}\email{philipp.blumer@cern.ch}

\author[1]{\fnm{Svenja} \sur{Geissmann}}

\author[2]{\fnm{Arnaldo J.} \sur{Vargas}}

\author[3]{\fnm{Gianluca} \sur{Janka}}

\author[4]{\fnm{Ben} \sur{Ohayon}}

\author*[1]{\fnm{Paolo} \sur{Crivelli}}\email{paolo.crivelli@cern.ch}

\affil[1]{\orgdiv{Institute for Particle Physics and Astrophysics}, \orgname{\textsc{Eth Zurich}}, \orgaddress{\city{Zurich}, \postcode{8093}, \country{Switzerland}}}

\affil[2]{\orgdiv{Laboratory of Theoretical Physics, Department of Physics}, \orgname{University of Puerto Rico}, \orgaddress{R\'io Piedras}, \postcode{00936},  \country{Puerto Rico}}

\affil[3]{PSI Center for Neutron and Muon Sciences CNM, 5232 Villigen PSI, Switzerland}

\affil[4]{\orgdiv{Physics Department}, \orgname{Technion—Israel Institute of Technology}, \orgaddress{Haifa}, \postcode{3200003}, \country{Israel}}

\abstract{We review the status of the QED calculations for the muonium $2S_{1/2}-2P_{3/2}$ energy interval and provide the updated theoretical value of $\SI{9874.357\pm0.001}{\MHz}$.
Additionally, we present a model for probing Lorentz-violating coefficients within the Standard Model Extension framework using the fine structure measurement in the presence and absence of a weak external magnetic field, enabling novel tests of CPT and Lorentz symmetry.
Using Monte Carlo simulations, we estimate that a precision of $\sim\SI{10}{\kilo\hertz}$ on the isolated $2S_{1/2}, F=1 - 2P_{3/2}, F=1$  transition could be achievable employing Ramsey's separate oscillatory fields (SOF) technique. Collecting the required statics will become feasible with the upcoming High-Intensity Muon Beam (HiMB) at the Paul Scherrer Institute (PSI) in Switzerland. These advancements will enable precise tests of radiative QED corrections and nuclear self-energy contributions, while also providing tests of new physics and sensitivity to unconstrained coefficients for Lorentz violation within the Standard Model Extension framework.
}


\maketitle

\section{Introduction}\label{intro}
Muonium (M), an exotic hydrogen-like atom formed by a positive muon (\posmu) and an electron (\electron), serves as a precision probe for bound-state quantum electrodynamics (QED) because its elementary constituents lack internal structure. This eliminates finite-size effects that often limit the sensitivity with which QED may be tested~\cite{2005_Karshenboim}.

Since its discovery in 1960 by V. Hughes, who observed its Larmor frequency in a magnetic field~\cite{1960_Hughes_M}, M has been subject to various studies using spectroscopic methods \cite{1999_Willmann,2000_Meyer,2022_Ohayon,2022_Janka} from which one can extract muon properties, test bound state QED, and probe potential new physics~\cite{2009_Jegerlehner,2021_Delaunay,2021_Balkin,2019_Fadeev, Frugiuele_2019,Stadnik:2022gaa}. 
Current experimental efforts aim to improve measurements of the $1S_{1/2}-2S_{1/2}$ energy splitting~\cite{2018_Crivelli_MuMASS}, the ground state hyperfine structure~\cite{2019_MuSEUM}, muonium to antimuonium conversion~\cite{2024_Bai}, or directly observe the gravitational interaction with elementary particles~\cite{Kaplan:2013fjr,2021_Soter}.

The Standard Model Extension (SME) offers a systematic framework for probing Lorentz and CPT symmetry~\cite{1997_Colladay,1998_Colladay}. This approach builds on the possibility that tiny deviations from Lorentz symmetry might be a low-energy signal for physics beyond the Standard Model and General Relativity~\cite{1989_Kostelecky,1991_Kostelecky,1995_Kostelecky}. The SME allows Lorentz-violating operators corresponding to different particle types to couple with distinct coefficients for Lorentz violation. Therefore, a systematic test of Lorentz and CPT symmetry requires considering experiments involving different systems, including second-generation particles such as muons bound in M atoms. 
M measurements have already been employed to constrain SME coefficients in the muonic sector through transitions such as the Zeeman-hyperfine transitions within the ground state~\cite{2001_Hughes,2000_Bluhm,2014_Gomes}, the $1S_{1/2}-2S_{1/2}$ transition~\cite{2014_Gomes}, and the classical Lamb shift $2S_{1/2}-2P_{1/2}$~\cite{2014_Gomes,2022_Ohayon}.

An area of particular interest is the M fine structure interval, known as the $2S_{1/2} - 2P_{3/2}$ energy transition. This transition allows for testing both QED and Lorentz and CPT symmetry, providing an opportunity to probe previously undetermined SME coefficients.
Despite its relevance in hydrogen-like atoms, it has been experimentally investigated only once, in 1990~\cite{1990_Kettell}.
This work revisits this transition by reviewing the latest theoretical higher-order QED calculations and updating the predicted values of the transition frequencies for the $2S_{1/2} - 2P_{3/2}$ and $2P_{1/2} - 2P_{3/2}$ intervals. Additionally, the potential to explore new SME coefficients is examined in two scenarios: one with a weak external magnetic field and another in the absence of such a field.

The experimental feasibility of measuring the fine structure transition is evaluated using a single interaction region and separated oscillatory fields (SOF) techniques. With detailed Monte-Carlo simulations validated with the measurements of the M Lamb shift by Mu-MASS ~\cite{2022_Ohayon,2020_Janka_2S}, we estimate the achievable precision using the existing Low-Energy Muon (LEM) beamline at the Paul Scherrer Institute (PSI). Furthermore, potential advancements in precision with the upcoming High-Intensity Muon Beam~\cite{2021_Aiba_HIMB} and muCool~\cite{2021_Antognini_muCool} are considered. 

\section{Calculation of the muonium fine-structure}\label{sec:fs_calc}
\begin{figure}[h]
    \centering
    \includegraphics[width=\linewidth]{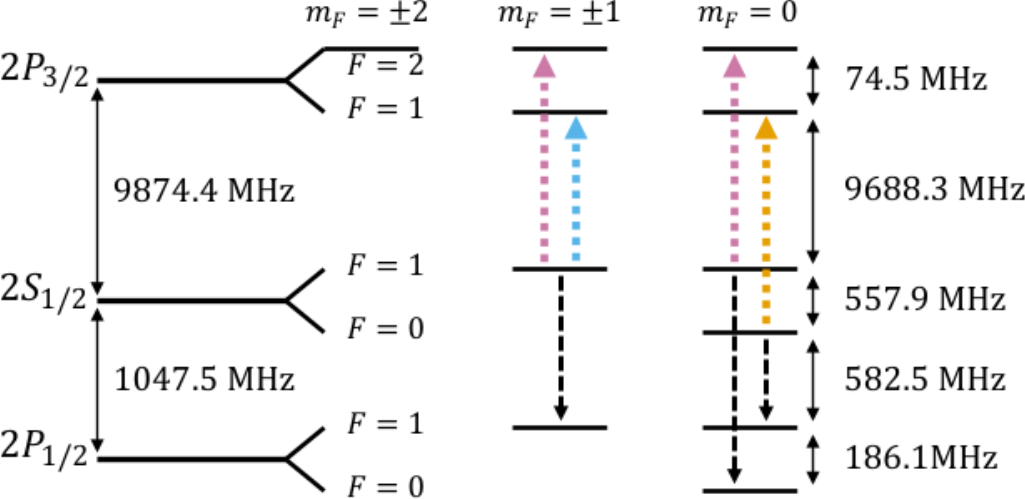}
    \caption{Diagram of the M energy levels including hyperfine structure for the $2S$ and $2P$ states. The atomic angular momentum $F$ projection along a specified axis is defined with the quantum number $m_F$. The arrows mark the allowed electric dipole transitions for the fine structure (colored dotted lines) and Lamb shift transitions (black dashed lines).}
    \label{fig:M_level}
\end{figure}

The energy value of the M $2S_{1/2} - 2P_{3/2}$ transition is based on well-established calculations for the hydrogen (H) atom which are summarized in Ref.~\cite{2022_CODATA, 2016_Hessels_Tab}. Higher-order QED corrections that are negligible in H but are relevant for the M $n=2$ manifold are described in Ref.~\cite{2022_Janka_EXA}. We further include recently calculated contributions from Ref.~\cite{2023_Adkins,2023_Eides}. The energy levels are characterized with $nL_J (F)$ using the principal quantum number $n$, the orbital angular moment $L$, and the total angular momentum $J=L+S$, where $S$ is the electron spin. The hyperfine structure due to the interaction between the electron and nuclear spin ($I$) is considered in the quantum number of the atomic angular momentum $F=J+I$. 
The energy state of interest $E_{nL_J}$ is the centroid of all hyperfine energy levels $E_{nL_JF}$, from which a radiative transition is allowed: 
\begin{equation}\label{eq:centroid}
     E_{nL_J} = \frac{\sum_F (2F + 1) E_{nL_JF}}{\sum_F(2F + 1)}.
\end{equation}
The energy eigenstates and transitions of the M $n=2$ manifold are visualized in Fig.~\ref{fig:M_level}.
In the following, we tabulate contributions that were not described in Ref.~\cite{2022_Janka_EXA} or have been updated since then and are relevant to the fine structure value in increasing order of $(Z\alpha)$, where $Z$ is the atomic number and $\alpha$ the fine structure constant.

The largest contribution to the fine structure interval comes from the Dirac eigenvalue $E_\text{M}$ for an electron of mass $m_e$ bound by a nucleus of finite mass $m_n$. We use the total mass $M=m_e + m_n$, the reduced mass $m_r=m_e m_n / (m_e+m_n)$, and the vacuum speed of light $c$. In contrast to Eq.~30 of CODATA 2022~\cite{2022_CODATA}, we define $E_\text{M}$ without the Barker-Glover recoil correction $E_\text{BKG}$ and thus write: 
\begin{equation}
\label{eq:EM}
    \begin{split}
        E_\text{M} = &Mc^2 + (f(n,\kappa)-1)m_r c^2 \\
        &- (f(n,\kappa)-1)^2 \frac{m_r^2 c^2}{2M}
    \end{split}
\end{equation}
where $f(n,\kappa) = (1 + \frac{(Z\alpha)^2}{(n-\delta)^2})^{-1/2}$, $\delta = |\kappa|-\sqrt{\kappa^2 - (Z\alpha)^2}$, and $\kappa = (L-J)(2J+1)$. 

Pure recoil contributions have been expanded in mass terms up to the order $(Z\alpha)^6$ in Ref.~\cite{2002_Blokland} and completed in Ref.~\cite{2023_Adkins} for the $S$ states. The updated calculations contain the previously established hard contributions and now include the newly added soft contributions in $E_\text{rec,R2}$:
\begin{equation}
\label{eq:R2}
    \begin{split}
        E_\text{rec,R2}(nS) = \frac{(Z\alpha)^6}{n^3}\biggl(\frac{m_r}{m_e}\biggr)^3\biggl(\frac{m_e}{m_n}\biggr)^2 \times \\
        \biggl[\underbrace{\frac{4}{\pi^2}\ln{\frac{m_n}{m_e}} - \frac{8}{3}\ln{\frac{m_n}{m_e}} - \frac{12\zeta_3}{\pi^2} + \frac{3}{\pi^2} + \frac{8}{3}}_\text{hard} \\
        \underbrace{-\frac{4}{9} -\frac{2}{n} - \frac{1}{3n^2} -\frac{1}{16n^3}}_\text{soft} \biggr] m_e c^2
    \end{split}
\end{equation}
Corrections for the $nP$ levels are derived in Ref.~\cite{2024_Patkos_Pstate}. We verified that they are negligible at our aimed precision.

Radiative recoils for the $nP$ states are addressed in Ref.~\cite{2024_Patkos_radCor} and are of the order of magnitude of $\SI{100}{\hertz}$ for M and $\SI{10}{\hertz}$ for H. They are not directly utilized here because of the different derivation compared to Ref.~\cite{2022_CODATA,2016_Hessels_Tab} and we take the order of magnitude contribution as its uncertainty.

Finally, we include the $(Z\alpha)$ expansion from the one-loop self-energy of the electron in the nuclear self-energy contribution $E_\text{NSE}$, as mentioned in Ref.~\cite{2023_Cortinovis}:
\begin{equation}
 \label{eq:esen}
    \begin{split}
    E_\text{NSE} &=\frac{ Z^2 \alpha (Z \alpha )^4}{ \pi n^3} \biggl(\frac{m_r}{m_e}\biggr)^3\biggl(\frac{m_e}{m_n}\biggr)^2 \times\\
    &\biggl[\biggl(\frac{10}{9} +\frac{4}{3}\ln\biggl[\frac{m_n}{m_r(Z \alpha )^2}\biggr]\biggr)\delta_{L0} \\
    &-\frac{4}{3}\ln(k_0) + (Z\alpha)A_{50}\biggr]m_e c^2
    \end{split}
\end{equation}
where $A_{50} = (\frac{139}{32} - 2\ln{2})\pi\delta_{L0}$.

Individual contributions are evaluated for the $2S_{1/2} - 2P_{3/2}$ and the $2P_{1/2} - 2P_{3/2}$ transitions of H and M and summarized in Tab.~\ref{tab:corr_overview}.
The constants from CODATA 2022~\cite{2022_CODATA} are used for the calculations, in particular, this means for the fine structure constant $\alpha$:
\begin{equation}
    \alpha = 1/137.035 999 177(21)
\end{equation} 
The propagation of uncertainty is performed as described in Ref.~\cite{2022_CODATA}, where each contribution is assigned a correlated ($u_0$) or an uncorrelated theoretical uncertainty ($u_n$). 
Table~\ref{tab:corr_overview} presents individual contributions for the $2S_{1/2} - 2P_{3/2}$ transition with uncertainties rounded to $\SI{1}{\kilo\hertz}$ and for the $2P_{1/2} - 2P_{3/2}$ transitions with uncertainties rounded to $\SI{0.1}{\kilo\hertz}$.
All contributions are included in the total sum and the uncertainty calculations.

In the case of M, the  $2S_{1/2} - 2P_{3/2}$ frequency transition is $\SI{9874.357\pm0.001}{\mega\hertz}$. Compared to hydrogen, relevant QED radiative corrections such as $E_\text{rel,R}$, $E_\text{RR}$, or nuclear self-energy $E_\text{NSE}$ can already be probed with precision measurements at the $\SI{10}{\kilo\hertz}$ level. 
The M hyperfine contribution for the $2P_{1/2} - 2P_{3/2}$ interval is $E_\text{HFS}=\SI{28.5}{\kilo \hertz}$. To experimentally extract this value, the fine structure measurement must be combined with the M Lamb shift measurement $2S_{1/2} - 2P_{1/2}$~\cite{2022_Ohayon}. 

\begin{table*}[ht]
\centering
\begin{tabular}{l|SS|SS}
\toprule
& \multicolumn{2}{c|}{$2S_{1/2} \rightarrow 2P_{3/2}$} & \multicolumn{2}{c}{$2P_{1/2} \rightarrow 2P_{3/2}$} \\
& \textbf{Hydrogen}  &\textbf{Muonium}  & \textbf{Hydrogen}  &\textbf{Muonium}  \\ 
& [MHz]  &[MHz]   & [MHz]  &[MHz]  \\ \midrule \midrule
$E_\text{M}$              & 10943.688 & 10896.947 & 10943.6881 & 10896.9473 \\

$E_\text{SE}$             & -1058.733   & -1045.761 & 25.3949 & 25.1784 \\ 
$E_\text{VP}$             & 26.853       & 26.510      & 0.0003 & 0.0003\\ 
$E_\text{VP, $\mu$+had}$  & 0.001        & 0.001       & & \\ 
$E_\text{2ph}$        & -0.104    & -0.103 & -0.0388 & -0.0385 \\ 
$E_\text{3ph}$        &  -0.000           & -0.000       &  0.0003             &  0.0003           \\
$E_\text{BG}$             & -0.001       & -0.084 & -0.0032 & -0.2524 \\ 
$E_\text{rel,S}$          & -0.358       & -3.138      &  & \\ 
$E_\text{rel,R}$       & 0.001        & 0.012  & -0.0000 & -0.0001 \\ 
$E_\text{rel,R2}$        & 0.000 & 0.001 & 0.0000 & 0.0000 \\
$E_\text{RR}$       & 0.002     & 0.014(1)    & 0.0000 & 0.0001(1)\\
$E_\text{Nuc}$            & -0.138    &             & -0.0000 & \\ 
$E_\text{NSE}$            & -0.001       & -0.041   & & \\[1ex]

$\biggl[E_\text{HFS}$            & 0.001         & 0.010      & 0.0028 & 0.0285$\biggr]$\\ [1ex]
\midrule \midrule
\makecell{\textbf{This work} \\ \textbf{without $E_\text{HFS}$}}    & 9911.209 & 9874.357(1) & 10969.0415 & 10921.8354(1) \\ 
Theory Ref.~\cite{2016_Hessels_Tab} & 9911.2093(3)    &   & 10969.0415  & \\
Theory Ref.~\cite{1990_Kettell}  & & 9874.3(3) &   & 10921.833(2)\\ 
\bottomrule 
\end{tabular}
\caption{Summary of the calculated contributions to the $2S_{1/2} - 2P_{3/2}$ and $2P_{1/2} - 2P_{3/2}$ transition in H and M. Uncertainties smaller than $\SI{1}{\kilo\hertz}$ and $\SI{0.1}{\kilo\hertz}$ for the two transitions are not listed, but are taken into account for the total uncertainty. Theoretical results from Ref.~\cite{2016_Hessels_Tab,1990_Kettell} are included for comparison.} \label{tab:corr_overview}
\end{table*}

\section{Prospects for testing Lorentz and CPT symmetry}\label{sec:SME}
Within the Standard Model Extension (SME), each Lorentz-violating operator couples to a coefficient referred to as a coefficient for Lorentz violation or SME coefficient. If Lorentz symmetry is exact, all such coefficients must be zero, and any nonzero coefficient would indicate a breaking of Lorentz symmetry. We examine the prospects for testing Lorentz and CPT symmetry by measuring the fine structure transition in the presence of a weak external magnetic field and subsequently discuss the scenario where such a field is absent.

\subsection{Lorentz-violating frequency shift in the presence of a weak magnetic field}

The structural similarity between M and H, both being two-fermion atoms with one fermion significantly heavier than the other, allows the results derived for H to be adapted to M. Specifically, the expressions for the Lorentz-violating energy shift of any atomic level in H, presented in subsections II.C and II.D of Ref.~\cite{2015_Kostelecky}, can be used to determine the Lorentz-violating frequency shift for the $2S_{1/2} - 2P_{3/2}$ transition in M.

The results in Ref.~\cite{2015_Kostelecky} are expressed in terms of nonrelativistic (NR) coefficients, which are linear combinations of the standard SME coefficients commonly appearing in the expressions for observables in nonrelativistic experiments and defined in detail in Ref.~\cite{2013_Kostelecky}. The key difference between the results in these subsections of Ref.~\cite{2015_Kostelecky} and those for M is the substitution of the proton NR coefficients in Eq.~(19) of that reference with the muon NR coefficients, 
\begin{eqnarray}
&&\TzBnrf{p}{kjm}\rightarrow \TzBnrf{\bar{\mu}}{kjm}, \nonumber \\
&&\ToBnrf{p}{kjm}\rightarrow \ToBnrf{\bar{\mu}}{kjm}, \nonumber\\
&&\Vnrf{p}{kjm}\rightarrow \Vnrf{\bar{\mu}}{kjm}.
\end{eqnarray}
These effective coefficients can be expressed in terms of SME muon NR coefficients with a definite CPT sign as follows
\begin{eqnarray}
 \TzBnrf{\bar{\mu}}{kjm}&=&-\gzBnrf{\mu}{kjm}-\HzBnrf{\mu}{kjm},\nonumber \\
 \ToBnrf{\bar{\mu}}{kjm}&=&-\goBnrf{\mu}{kjm}-\HoBnrf{\mu}{kjm},\nonumber \\
\Vnrf{\bar{\mu}}{kjm}&=& \cnrf{\mu}{kjm}+\anrf{\mu}{kjm}.
\end{eqnarray}

The index $\bar{\mu}$ on the left side of the equation indicates that these effective coefficients are associated with the antimuon, while the index $\mu$ on the right side denotes the muon SME coefficients. The SME coefficients on the right-hand side are related to Lorentz-violating operators with a definite CPT sign: the $g$-type and $a$-type coefficients are associated with CPT-odd operators, while the $H$-type and $c$-type coefficients correspond to CPT-even operators. The distinction between the ${\mathcal V}$-type and ${\mathcal T}$-type coefficients for the muon and antimuon lies in the sign in front of the coefficients associated with CPT-odd operators, specifically the $a$-type and $g$-type coefficients. 

With these considerations, we can use the general results presented in Ref.~\cite{2015_Kostelecky} to obtain the Lorentz-violating energy shift for the energy states $2S_{1/2}$ and $2P_{3/2}$ in the presence of a weak magnetic field, meaning a field that produces a Zeeman shift smaller than the dominant contribution to the hyperfine structure of M. Finally, the Lorentz-violating frequency shift is the difference between these energy shifts.  

We use the same definition for the relevant quantum numbers as in Sec.~\ref{sec:fs_calc} and specify the quantum number $m$ for the projection of $\vec{F}$ along the direction of the applied magnetic field.

Using the general expressions for the energy shift presented in Ref.~\cite{2015_Kostelecky}, we find that the Lorentz-violating energy shift $\delta \epsilon_{1/2}^{(Fm)}$ for the $2S_{1/2}^{(Fm)}$ level in the presence of a weak magnetic field is given by
\begin{equation}
    \begin{split}
        \delta \epsilon^{(Fm)}_{1/2}=-\dfrac{m}{2 \sqrt{3 \pi}}\sum_{k=0,2,4}\langle|{\boldsymbol p}^k|\rangle_{20}\biggl(\TzBnrf{\bar{\mu}}{k10} \\
        +2\ToBnrf{\bar{\mu}}{k10}\biggr) -\dfrac{1}{\sqrt{4\pi}} \Vnrf{\bar{\mu}}{400}\langle |{\boldsymbol p}^4|\rangle_{20}.
    \end{split}
\label{S1/2}
\end{equation}
Here, we consider only the contributions from the muon coefficients and ignore all contributions from the electron coefficients, as ordinary-matter experiments are more sensitive to them \cite{2000_Bluhm, 2014_Gomes}. Additionally, we disregard contributions from coefficients that affect the Lorentz-violating energy shift for both $2S_{1/2}$ and $2P_{3/2}$ in the same way, as these contributions will cancel out when calculating the Lorentz-violating frequency shift. 

The momentum expectation values that contribute to the frequency shift are given by
\begin{eqnarray}
\langle |{\boldsymbol p}^0|\rangle_{nL}&=& 1,
\nonumber\\
\langle |{\boldsymbol p}^2|\rangle_{nL}&=& \left(\dfrac{\alpha m_r}{n}\right)^2,
\nonumber\\
\langle |{\boldsymbol p}^4|\rangle_{nL}&=&
\left(\dfrac{\alpha m_r}{n}\right)^4
\left(\dfrac{8n}{2L+1}-3\right).
\label{radialExp}
\end{eqnarray}

The Lorentz-violating energy shift  $\delta \epsilon_{3/2}^{(Fm)}$  for the $2P_{3/2}^{(Fm)}$ energy state in the presence of a weak magnetic field is given by
\begin{eqnarray}
\delta \epsilon_{3/2}^{(Fm)}&=&-\sum_{j=1,3}\sum_{k=0,2,4}\beta^{(j)}_{(F,m)}\langle |{\boldsymbol p}^k|\rangle_{21}\TzBnrf{\bar{\mu}}{kj0}\nonumber\\
&&-\sum_{j=1,3}\sum_{k=0,2,4}\alpha^{(j)}_{(F,m)}\langle |{\boldsymbol p}^k|\rangle_{21}\ToBnrf{\bar{\mu}}{kj0}\nonumber\\
&&- \dfrac{ (2F-5)(F(F+1)-3m^2)}{12\sqrt{5 \pi}}\sum_{k=2,4}\langle |{\boldsymbol p}^k|\rangle_{21}\Vnrf{\bar{\mu}}{k20}\nonumber\\
&&-\dfrac{1}{\sqrt{4\pi}} \Vnrf{\bar{\mu}}{400}\langle |{\boldsymbol p}^4|\rangle_{21}.
\label{P3/2}
\end{eqnarray}
Here, the angular expectation values $\beta^{(j)}_{(F,m)}$ are given by
\begin{eqnarray}
\beta^{(1)}_{(1,m)}&=&\dfrac{m}{4 \sqrt{3 \pi}},\nonumber \\
\beta^{(1)}_{(2,m)} &=& \dfrac{3m}{20 \sqrt{3 \pi}},\nonumber\\
\beta^{(3)}_{(1,m)} &=& 0,\nonumber\\
\beta^{(3)}_{(2,m)} &=& \dfrac{m(17-5 m^2)}{20 \sqrt{7 \pi}},
\end{eqnarray}
and the ones for $\alpha^{(j)}_{(F,m)}$ by
\begin{eqnarray}
\alpha^{(1)}_{(F,m)} &=&-4\beta^{(1)}_{(F,m)}, \nonumber\\
 \alpha^{(3)}_{(F,m)} &=& -\sqrt{\dfrac{8}{3}}\beta^{(3)}_{(F,m)}.
\end{eqnarray}
At last, the Lorentz-violating frequency shift for the transition $2S_{1/2}^{(F'm')} - 2P_{3/2}^{(Fm)}$ is
\begin{equation}
2\pi \delta \nu=\delta \epsilon_{3/2}^{(Fm)}-\delta \epsilon_{1/2}^{(F'm')}.
\label{2S-2P}
\end{equation}

The $a$-type and $c$-type muon coefficients with $j=2$ remain unconstrained~\cite{2011_Kostelecky_Table}. The primary advantage of studying the $2S_{1/2} - 2P_{3/2}$ transition, compared to the $2S_{1/2} - 2P_{1/2}$ transition in the context of the SME, is that the former is sensitive to these unconstrained coefficients as the ${\mathcal V}$-type coefficients, also known as spin-independent coefficients, with index $j$ can only contribute to the energy shift of states where $(j+1)/2 \leq J$, as shown in Ref.~\cite{2015_Kostelecky}. Consequently, the coefficients with $j=2$ cannot contribute to the energy shift of $2P_{1/2}$ but contribute to the one for $2P_{3/2}$.

If the transition involves states in the hyperfine sublevel $F=2$ of the $2P_{3/2}$ state, the transition frequency will also be sensitive to the ${\mathcal T}$-type coefficients, also known as spin-dependent coefficients, with $j=3$. Note that for $F=1$, the corresponding angular expectation values $\beta^{(3)}_{(1,m)}$ and $\alpha^{(3)}_{(1,m)}$ vanish. They vanish because spin-dependent coefficients with index $j$ can only contribute to the energy shift of states with $(j+1)/2 \leq F$ \cite{2015_Kostelecky}. Therefore, the $j=3$ coefficients can only contribute to states with $2 \leq F$.

The frequency shift~\eqref{2S-2P} was obtained in the laboratory frame. Since the SME coefficients are frame-dependent, every bound on the coefficients must be reported in the same reference frame to allow for meaningful comparisons. The frame commonly used for this is the Sun-centered frame~\cite{2002_Kostelecky_sunframe}. The details of how to transform the coefficients from a laboratory on the surface of the Earth to the Sun-centered frame have been thoroughly discussed in Ref.~\cite{2015_Kostelecky}. Following the procedure described in that reference, we obtained that the form of the Lorentz-violating frequency shift takes the form
\begin{equation}
\label{lvlab}
    \begin{split}
        \delta \nu= A_0 +\sum_{m=1}^3 \biggl(&A^{(s)}_m \sin{(m\,\omega_\oplus T_L)} \\
        &+A^{(c)}_m \cos{(m\,\omega_\oplus T_L)}\biggr),
    \end{split}
\end{equation}
where $\omega_\oplus = 2\pi / (23{\rm h}\, 56{\rm m})$ is the sidereal frequency, and $T_L$ is a convenient offset of the time coordinate $T$ in the Sun-centered frame \cite{2015_Kostelecky}.

The most prominent signal of Lorentz violation from \eqref{lvlab} is a sidereal variation of the resonance frequency. This variation occurs because, in the presence of Lorentz violation, the spectrum of M may depend on its overall orientation. Therefore, a sidereal variation of the resonance frequency may be observed as the system, M with the applied magnetic field, rotates with the Earth relative to the Sun-centered frame.

In this discussion, however, we focus on the constant frequency shift $A_0$ as it allows for discrepancies between experimental results and Standard Model predictions, as shown in previous works \cite{2014_Gomes, 2015_Kostelecky}. To isolate this effect, we ignore all time-dependent contributions and concentrate solely on the constant shift. Accordingly, we take the Lorentz-violating frequency shift as $ \delta \nu_0  =A_0$, where $\delta \nu_0$ is obtained using the following replacements in \eqref{2S-2P}

\begin{eqnarray}
 \Vnrf{\bar{\mu}}{k00} &\rightarrow&  \sVnrf{\bar{\mu}}{k00}, \nonumber\\
 \Vnrf{\bar{\mu}}{k20} &\rightarrow&  \dfrac{1}{4}(1+3\cos{2\vartheta})\sVnrf{\bar{\mu}}{k20}, \nonumber\\
 \TzBnrf{\bar{\mu}}{k10}&\rightarrow& \cos{\vartheta}\sTzBnrf{\bar{\mu}}{k10},\nonumber\\
 \ToBnrf{\bar{\mu}}{k10}&\rightarrow&\cos{\vartheta} \sToBnrf{\bar{\mu}}{k10},\nonumber\\
  \TzBnrf{\bar{\mu}}{k30}&\rightarrow& \dfrac{1}{8} \left(3 \cos{\vartheta} +5 \cos{3 \vartheta}\right) \sTzBnrf{\bar{\mu}}{k30},\nonumber\\
 \ToBnrf{\bar{\mu}}{k30}&\rightarrow& \dfrac{1}{8} \left(3 \cos{\vartheta} +5 \cos{3 \vartheta}\right) \sToBnrf{\bar{\mu}}{k30},\nonumber\\
\end{eqnarray}
where the superscript ${\rm Sun}$ identifies the coefficients as those in the Sun-centered frame. The angle $\vartheta$ represents the angle between the applied magnetic field and Earth's rotation axis.  

As previously demonstrated in Ref.~\cite{2014_Gomes,2015_Kostelecky}, we can constrain the coefficients contributing to $A_0$ by assuming that any discrepancy between the Standard Model prediction $\nu_{\rm SM}$ and the experimental value $\nu_{\rm exp}$ is due to the presence of the constant shift $A_0$. Explicitly, we can constraint the coefficients by using the expression 
\begin{equation}
\nu_{\rm exp}-\nu_{\rm SM}=A_0.
\end{equation}

\subsection{Lorentz-violating frequency shift in a zero magnetic field}

The paradigm of testing Lorentz and CPT symmetry in spectroscopy experiments without the presence of external magnetic fields is discussed in subsection II.C of Ref.~\cite{2014_Gomes} and III.A of Ref.~\cite{2015_Kostelecky}. In the absence of a magnetic field and under the assumption of rotational symmetry, the energy levels exhibit a $(2F+1)$-fold degeneracy associated with the orientation of the atom in space, where $\vec{F}$ is the total angular momentum of the atom. This degeneracy indicates that the atomic spectrum is independent of its overall orientation, which is consistent with rotational symmetry. However, in the case of rotational symmetry breaking, the energy becomes orientation-dependent and the $(2F+1)$-fold degeneracy is lifted.

We will focus on the $2S_{1/2} - 2P_{3/2}^{(F=1)}$ transition, allowing us to disregard contributions from coefficients with $j=3$ as they only contribute to the $F=2$ case. Furthermore, we will exclude contributions from the $j=1$ coefficients as measurements of the Zeeman-hyperfine transition within the ground state of M are more sensitive to these coefficients~\cite{2014_Gomes}. This narrower scope, disregarding coefficients, is for convenience, as the expressions for the energy shift are significantly more intricate in the absence of an external magnetic field than when one is present.

Following the procedures described in subsection II.C of Ref.~\cite{2014_Gomes} and III.A of Ref.~\cite{2015_Kostelecky}, we can obtain the energy shift for the relevant energy states. The correction for the $2S_{1/2}^{F}$ is the same as for $F=0$ and $F=1$, since the only ${\mathcal V}$-type coefficients with $j=0$, called isotropic coefficients, can contribute. We get that 
\begin{equation} 
\delta \epsilon_{1/2}=-\dfrac{1}{\sqrt{4\pi}} \sVnrf{\bar{\mu}}{400}\langle |{\boldsymbol p}^4|\rangle_{20}. 
\end{equation} 
As before, we only consider contributions from the muon coefficients and disregard contributions from coefficients that affect the Lorentz-violating energy shift for both $2S_{1/2}$ and $2P_{3/2}$ in the same way.

The 3-fold degeneracy of the state $2P_{3/2}^{F=1}$ is lifted due to a breaking of rotational symmetry. We can use the label $\xi$, which takes values $-1$, $0$, and $1$, to denote the three new energy states that result from the splitting of the energy level $2P_{3/2}^{F=1}$ in the presence of rotational symmetry breaking. The Lorentz-violating energy shift, using the notation introduced in Ref.~\cite{2014_Gomes}, is given by
\begin{equation}
\delta \epsilon^{(1,\xi)}_{3/2}=-\dfrac{1}{\sqrt{4\pi}} \sVnrf{\bar{\mu}}{400}\langle |{\boldsymbol p}^4|\rangle_{21}+u_\xi D+u_\xi^* \dfrac{\Delta_0}{D}
\label{nomag}
\end{equation}
where
\begin{equation}
u_\xi=\dfrac{1+i\sqrt{3}\xi}{3(1-3\xi^2)}
\end{equation}
and
\begin{equation}
D=\left(\dfrac{\Delta_1+\sqrt{\Delta_1^2-4 \Delta_0^3}}{2}\right)^{1/3}.
\end{equation}
The parameters $\Delta_0$ and $\Delta_1$ represent rotational scalars formed from the coefficients of Lorentz violation. The explicit expressions are given by
\begin{eqnarray}
\Delta_0&=&\dfrac{9 \sqrt{5}}{80 \pi} \sum_{m=-2}^2 \langle 2m2(-m)|00\rangle A_m A_{-m}.\nonumber\\
\Delta_1&=&\dfrac{189}{160 \sqrt{14} \pi^{3/2}} \sum_{s=-2}^2 \sum_{m=-2}^2\sum_{q=-2}^2\langle 2m2q2(m+q)\rangle\nonumber\\
&&\times\langle 2s2(q+m)|00\rangle A_m A_q A_{s},
\label{S1S2}
\end{eqnarray}
where $\langle j_1m_1j_2m_2|j_3m_3\rangle$ are the Clebsch-Gordan coefficients.  The amplitudes $A_m$ are defined as
\begin{equation}
A_m=-\sum_{k=2,4} \langle |{\boldsymbol p}^k|\rangle_{21}\sVnrf{\bar{\mu}}{k2m},
\label{Am}
\end{equation}
and they play the same role as the $q_{2m}$ coefficients defined in Ref.~\cite{2014_Gomes}, with the difference being that the combination of coefficients contributing to $A_m$ are those for antimuons, rather than muons, which is the case for $q_{2m}$.

The energy shift~\eqref{nomag} is expressed in the Sun-centered frame at the zeroth-boost order. By the zeroth-boost order approximation, we mean to approximate the transformation from the Earth-based laboratory frame to the Sun-centered frame as a pure rotation~\cite{2015_Kostelecky}. The terms contributing to~\eqref{nomag} are combinations of coefficients for Lorentz violation that form rotational scalars, as discussed in detail in Ref.~\cite{2014_Gomes} and Ref.~\cite{2015_Kostelecky}. In particular, the parameters $\Delta_0$ and $\Delta_1$ have the same value in any two frames rotated relative to each other. Their invariance under rotations allows us to express $A_m$ in~\eqref{Am} in terms of the coefficients in the Sun-centered frame at zeroth-boost order, as it will have the same form in this frame as in the laboratory frame. Therefore, the energy shift \eqref{nomag} is constant in contrast to the situation where an external magnetic field is present, causing the energy shift to change with Earth's rotation.

The sidereal variation seen in the presence of an external magnetic field arises because the energy shift depends on the relative orientation between the Lorentz-violating background fields and the magnetic field. The dominant contribution owing to Lorentz violation in this case is due to the projection of the Lorentz-violating fields along the direction of the magnetic field. As the magnetic field rotates with Earth, the components of the Lorentz-violating background field in a fixed inertial reference frame, such as the Sun-centered frame, that dominate the Lorentz-violating energy shift will change, giving rise to the sidereal variation.

The signal for Lorentz violation in the case of no external magnetic field is different. We can begin the discussion by defining the Lorentz-violating frequency shift for the transition $2S_{1/2}^F - 2P_{3/2}^{(F=1,\xi)}$ as
\begin{equation}
2\pi \delta \nu = \delta \epsilon^{(1,\xi)}_{3/2} - \delta \epsilon_{1/2}. 
\end{equation}
From this frequency shift, we can deduce that instead of observing only one resonance peak corresponding to the hyperfine level $F=1$ in the state $2P_{3/2}$, we should observe two or three resonance peaks very close to each other. The reason there can be two or three is that if
\begin{equation}
 \Delta_1^2-4 \Delta_0^3 = 0, 
\end{equation}
then the energy levels corresponding to $\xi=-1$ and $\xi=1$ are degenerate, and we will observe only two peaks. Otherwise, we will have three distinct peaks.

The strategy for imposing a bound on Lorentz violation should rely on the presence or absence of multiple unexpected peaks in the data. If the analysis concludes, within the uncertainty level, that there is no evidence of multiple peaks, any contribution of the terms in~\eqref{nomag} that depend on $\xi$ should be within the uncertainty of the analysis. The remaining contribution to the transition frequency is
\begin{equation}
 2\pi \delta \nu = -\dfrac{1}{\sqrt{4\pi}} \sVnrf{\bar{\mu}}{400}(\langle |{\boldsymbol p}^4|\rangle_{21} - \langle |{\boldsymbol p}^4|\rangle_{20}), 
\end{equation}
which can be constrained by comparing theoretical and experimental values for the Lamb shift, as previously done~\cite{2014_Gomes, 2022_Ohayon}.

Another possibility, as discussed in Ref.~\cite{2014_Gomes}, is to assume that the presence of multiple resonances might broaden the lineshape for the transition $2S_{1/2}^F - 2P_{3/2}^{(F=1)}$. For example, if we assume that all energy sublevels, $\xi = -1, 0, 1$, were excited during the experiment with equal probability, this would contribute an amount
\begin{equation} 
\label{eq:sme_width}
\sigma_\nu^2 = \dfrac{2}{9} \Delta_0 
\end{equation}
to the square of the statistical uncertainty $\sigma_\nu$ of the resonance frequency. A more rigorous analysis in this direction could, in principle, be used to impose bounds on the coefficients for Lorentz violation.
%
%
%
%
\section{Microwave spectroscopy of the M $2S_{1/2} - 2P_{3/2}$ energy transition}\label{mw_spec}

So far, the M experimental measurements agree well with the theoretical QED calculations~\cite{2000_Meyer,2022_Ohayon,2022_Janka,2001_Hughes,2000_Bluhm,2014_Gomes}. However, they are all currently limited by statistics. Leveraging advances of the High-Intensity Muon Beam (HiMB) at PSI in Switzerland~\cite{2021_Aiba_HIMB} together with muCool~\cite{2021_Antognini_muCool} will allow reaching the sub-MHz precision level for the $1S_{1/2}-2S_{1/2}$~\cite{2023_Cortinovis} and $2S_{1/2}-2P_{1/2}$~\cite{2022_Janka_EXA} measurements conducted by the Mu-MASS collaboration. In the following, we explore a precision measurement of the $2S_{1/2}-2P_{3/2}$ fine structure transition.

The short lifetime of M ($\tau_\text{M} = \SI{2.2}{\micro\second}$) requires a compact experimental setup with the drawback that contamination from long-lived excited states can distort the lineshape. The Lamb shift measurement with individual hyperfine transitions ranging from $\SI{558}{\mega\hertz}$ to $\SI{1326}{\mega\hertz}$~\cite{2022_Ohayon,2022_Janka} is affected by the $3S_{1/2}-3P_{1/2}$ or $4S_{1/2}-4P_{3/2}$ intervals with frequencies around $312\MHz$ and $1233\MHz$, respectively. The advantage of the $2S_{1/2}-2P_{3/2}$ interval compared to the $2S_{1/2}-2P_{1/2}$, is that it is isolated from such contamination.

\begin{figure}
    \centering
    \includegraphics[width=\linewidth]{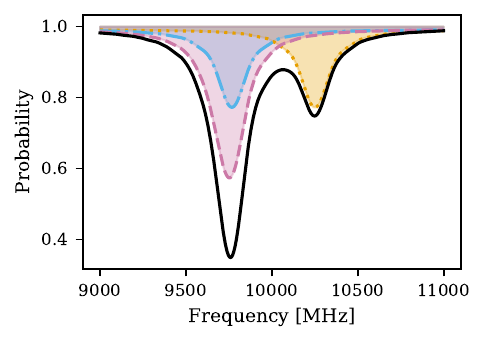}
    \caption{The dotted (yellow), dash-dotted (blue), and dashed (magenta) lines are an example of the numerical solutions of the frequency transitions with the initial states $2S_{1/2}, F=0$, $2S_{1/2}, F=1, m_F=0$ and $2S_{1/2}, F=1,m_F=\pm1$, respectively. The solid (black) lineshape corresponds to the measured fine structure lineshape assuming that the initial $2S$ states are populated evenly.}
    \label{fig:fs-lineshape}
\end{figure}

We consider the simplest experimental setup without the presence of an external magnetic field, where the atoms pass through a microwave field with electric field amplitude $E_0$, frequency $\omega$, and interaction time $T$, allowing the atoms initially in the $2S$ state to transition to the $2P$ state. The transition probability is calculated by numerically solving the according Bloch equations~\cite{2018_Marsman_BEq}. Once the atom is in the $2P$ state, it relaxes to the ground state with a radiative lifetime of $\tau_{2P}=\SI{1.6}{\ns}$. The remaining $2S$ atom population is measured, resulting in the depopulation signal as a function of the frequency. An example of the resonance lineshape for the fine structure measurement is shown in Fig.~\ref{fig:fs-lineshape}, with $\SI{10}{\keV}$ M, an electric field $E_0=\SI{25}{\volt/\cm}$ and an exposure time $T=\SI{7}{\ns}$ to the microwave field.

To accurately determine the center of the resonance line, addressing the broadening mechanisms that distort this position is essential. The natural linewidth of the individual transitions is given by $\Delta \nu = 1/(2\pi\tau_{2P}) = 99.7\MHz$. However, it is broadened in practice because of different experimental configurations such as the atom's kinetic energy distribution or microwave field inhomogeneities. As shown in Fig.~\ref{fig:fs-lineshape}, overlapping transition lines further distort the resonance center. 
To overcome these limitations, Ramsey's separate oscillatory fields (SOF) spectroscopy~\cite{1949_Ramsey} can significantly reduce the effective resonance width. We focus on the isolated $2S_{1/2}, F=0 - 2P_{3/2}, F=1$ transition (yellow in Figs.~\ref{fig:M_level} and \ref{fig:fs-lineshape}) and exploit the interference effects between two spatially separated microwave fields.

In the SOF method, the atoms pass through two microwave interaction regions separated by a field-free region, where the atomic states evolve freely during the time interval $\tau$. The two microwave fields are applied either in-phase ($P_0^\text{SOF}$) or with a relative phase shift $\pi$ ($P_\pi^\text{SOF}$). Subtracting the measured remaining $2S$ population for these two configurations, $P_\pi^\text{SOF} - P_0^\text{SOF}$, produces an interference pattern with a significantly narrower central fringe compared to the single field lineshape. 

\begin{figure*}
    \centering
    \includegraphics[width=\linewidth]{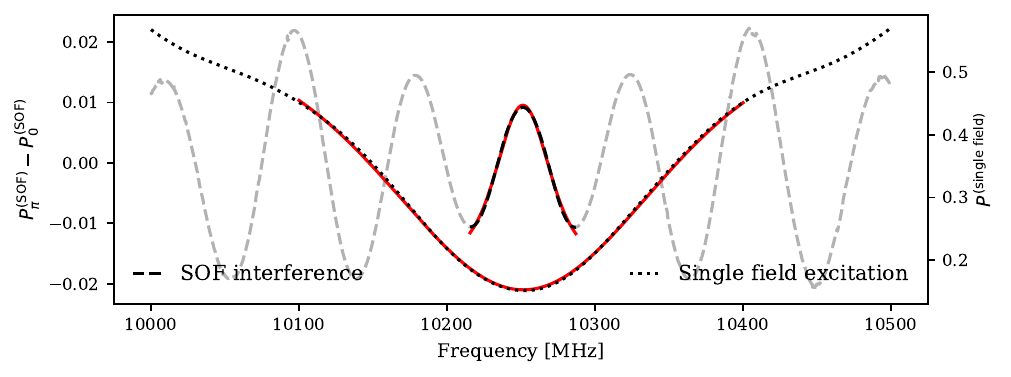}
    \caption{Numerical solution of the 2-level Bloch equations for single field and SOF spectroscopy considering M with $10\keV$ kinetic energy traversing microwave cavities and field free region of $\SI{2}{\cm}$ length and electric field amplitude of $\SI{60}{\volt/\cm}$. }
    \label{fig:rabi-ramsey-numerical}
\end{figure*}

\begin{figure}[h]
    \centering
    \includegraphics[width=\linewidth]{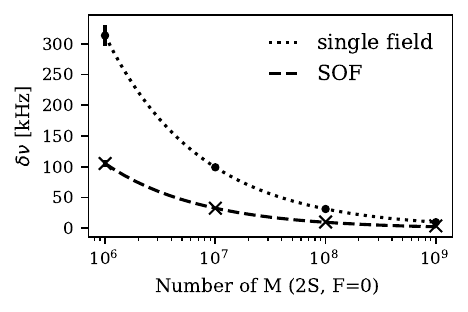}
    \caption{The precision of the resonance center $\delta \nu$ for single field (dot) and SOF (x) spectroscopy is shown depending on the number of initial $2S, F=0$ states. Fitting an inverse square-root function yields the relation between the precision $\delta\nu$ and the statistics $N$.}
    \label{fig:precision}
\end{figure}

Fig.~\ref{fig:rabi-ramsey-numerical} shows the numerical solutions for both SOF and single field M spectroscopy at kinetic energy $E_\text{kin} = 10\keV$. The simulation considers atoms traversing the microwave and field-free regions of length $\SI{20}{\mm}$ under an electric field amplitude $E_0 = \SI{60}{\volt/\cm}$. The resonance center is determined by fitting a Voigt function (red) to the resonances, representing the convolution of Lorentzian and Gaussian distributions.
Based on these numerical solutions, the precision of the resonance center can be determined under varying statistical conditions. 
The total amount of M atoms, $N=\sum_i n_i$, in the $2S_{1/2}, F=0$ state is evenly distributed for each measured frequency point $i$. Each $n_i$ is then randomized according to a Gaussian distribution around its mean value, $n_i \pm \sqrt{n_i}$, and scaled according to the calculated transition probability.
For SOF spectroscopy, the total number of atoms $N$ is divided equally between the two configurations ($P_{0}^\text{SOF}$, $P_\pi^\text{SOF}$). Fitting the randomized resonance lineshape of the single field and SOF central fringe with a Voigt function determines the resonance center and its precision $\delta \nu$. The mean precision and its standard deviation from 1000 fits of randomized distributions per statistical configuration $N$ are shown for single field interaction (dots) and SOF spectroscopy (x) in Fig.~\ref{fig:precision}. The fitted lines show the inverse square-root relation between precision $\delta\nu$ and statistics $N$.

The HiMB project aims to increase the surface muon rate at PSI to $10^{10}\,\mathrm{Hz}$ by improving the proton target and muon beamline~\cite{2021_Aiba_HIMB}. Together with muCool~\cite{2021_Antognini_muCool} a keV muon beam could be produced with a sufficiently small phase space that would allow the use of a gas cell as a M converter, similar to what has been done in the most precise H Lamb shift measurement that reached a precision of $\SI{3}{\kilo \hertz}$~\cite{2019_Bezginov}. The combination of HiMB and muCool is expected to produce a M rate of $5\times 10^5\,\mathrm{Hz}$~\cite{2021_Antognini_muCool}.

Considering a detection efficiency of $16\%$~\cite{2022_JankaPhD} and that about $\SI{10\pm3}{\%}$ of the M atoms are produced in the initial $2S$ state~\cite{2020_Janka_2S} of which $1/4$ populate the $2S, F=0$ state, we find that a statistical precision of the fine structure of M at the $\sim\SI{10}{\kilo\hertz}$ level is feasible within 10 days of beamtime.

\section{Conclusions}\label{conclusion}
The transition frequency of the M fine structure $2S_{1/2} - 2P_{3/2}$ has been calculated as $\SI{9874.357\pm0.001}{\MHz}$. Because of the smaller nuclear mass of M, radiative QED and nuclear self-energy corrections can already be probed with a precision measurement at the $\sim\SI{10}{\kilo\hertz}$ level, compared to H, which is sensitive only in the $\sim\SI{1}{\kilo\hertz}$ regime.
Additionally, we developed models to test Lorentz and CPT symmetry using fine structure measurements, both with and without an external magnetic field. These models show that there are coefficients to which a fine structure measurement in M is uniquely sensitive.

Due to the isolated energy eigenstates compared to the Lamb shift measurement, contamination from long-lived excited states is negligible~\cite{2022_Ohayon,2022_Janka}, a key advantage for precision studies. Using a Monte Carlo simulation, the expected precisions of a single interaction zone and SOF-type measurements are studied for the isolated $2S_{1/2}, F=0 - 2P_{3/2}, F=1$ frequency transition in the absence of an external magnetic field. Combining the upcoming HiMB beamline~\cite{2021_Aiba_HIMB} together with the muCool apparatus~\cite{2021_Antognini_muCool} will allow to reach a precision of $\sim\SI{10}{\kilo\hertz}$.

\backmatter

\subsection*{Acknowledgments}
This work is supported by The Swiss National Science Foundation under grants 197346 and 219485, and ISF grant no. 2071390.

\subsection*{Author contributions}
All authors contributed to the conception and design of the study. SG and PB evaluated the calculations for the fine structure frequency transition values, as well as the simulations and statistical estimations. AJV calculated the SME contributions and authored the corresponding section.
All authors commented on previous versions of the manuscript. All authors read and approved the final manuscript.

\subsection*{Data Availability Statement}
The datasets generated during and/or analyzed during the current study are available from the corresponding author.

\bibliography{sn-bibliography}

\end{document}